\documentclass[a4paper]{article}

\usepackage[utf8]{inputenc}
\usepackage{tikz,booktabs,float,color,adjustbox,multirow,tabularx,cite}

\usepackage{INTERSPEECH2020}

\title{Streaming end-to-end bilingual ASR systems with joint language identification}
\name{Surabhi Punjabi\thanks{$^*$Equal contribution. $^\dagger$Work done during an internship at Amazon.}$^*$, Harish Arsikere$^*$, Zeynab Raeesy, Chander Chandak, Nikhil Bhave,\\Ankish Bansal$^\dagger$, Markus M{\"u}ller, Sergio Murillo, Ariya Rastrow, Sri Garimella,\\Roland Maas, Mat Hans, Athanasios Mouchtaris, Siegfried Kunzmann}
\address{Alexa Machine Learning, Amazon}
\email{\{spunjabi,arsikere,raeesyzr\}@amazon.com}

\begin{document}
\frenchspacing
\maketitle
\begin{abstract}
Multilingual ASR technology simplifies model training and deployment, but its accuracy is known to depend on the availability of language information at runtime. Since language identity is seldom known beforehand in real-world scenarios, it must be inferred on-the-fly with minimum latency. Furthermore, in voice-activated smart assistant systems, language identity is also required for downstream processing of ASR output. In this paper, we introduce streaming, end-to-end, bilingual systems that perform both ASR and language identification (LID) using the recurrent neural network transducer (RNN-T) architecture. On the input side, embeddings from pretrained acoustic-only LID classifiers are used to guide RNN-T training and inference, while on the output side, language targets are jointly modeled with ASR targets. 
The proposed method is applied to two language pairs: English-Spanish as spoken in the United States, and English-Hindi as spoken in India. Experiments show that for English-Spanish, the bilingual joint ASR-LID architecture matches monolingual ASR and acoustic-only LID accuracies. For the more challenging (owing to within-utterance code switching) case of English-Hindi, English ASR and LID metrics show degradation. Overall, in scenarios where users switch dynamically between languages, the proposed architecture offers a promising simplification over running multiple monolingual ASR models and an LID classifier in parallel.
\end{abstract}

\noindent\textbf{Index Terms}: multilingual, streaming speech recognition, language identification, end-to-end, RNN-T 

\section{Introduction}\label{sec:intro}

Multilingual automatic speech recognition (ASR) is an active area of research for two reasons: 
(1) it improves performance across languages, particularly the under-resourced ones, by allowing data sharing and cross-lingual knowledge transfer \cite{schultz2001language,ghoshalmultilingual,VeselyKGJE12,scanzio2008use,besacier2014automatic,nguyen2014multilingual}; and (2) it allows the same model to be used by two or more languages~\cite{vu2012first}, thereby creating seamless multilingual experiences via simplified model training, maintenance, and deployment.
In hybrid ASR systems with separate acoustic model (AM) and language model (LM) components, shared hidden layer training~\cite{Thomas,huang2013cross,arsikere2019multi} and transfer learning~\cite{heigold2013multilingual,feng2018improving,stuker2014training} have been proposed to address the data scarcity problem for under-resourced languages. However, achieving the model unification objective is more challenging in hybrid systems given the AM-LM separation. With recent end-to-end ASR advancements such as listen, attend and spell (LAS)~\cite{chan2015listen} and recurrent neural network transducers (RNN-T)~\cite{graves2012sequence}, unified multilingual ASR systems are now possible.
Among these popular frameworks, RNN-T is favored in online applications owing to its support for streaming ASR~\cite{he2019streaming}. We mainly focus on the objective of delivering a seamless experience to multilingual users without them having to specify the language before each interaction.
 
A commonly adopted architecture for creating a live multilingual experience is to run parallel ASRs behind the scenes while a language detector identifies spoken language~\cite{wang2019signal,chandak2020streaming}. Multilingual ASR systems can simplify such architectures by eliminating the need to run multiple monolingual ASRs, but previous research on multilingual end-to-end models suggests that language information (assumed to be known beforehand in most studies) plays a crucial role in achieving acceptable levels of performance~\cite{seki2018end,muller2018neural,kannan2019large,li2018multi,li2019bytes}.
While language information can be provided at runtime by preselecting the language, such an experience could be counter-intuitive because multilingual users often prefer to seamlessly switch between languages.
 
Streaming language detectors have been proposed as a solution for guiding multilingual ASR on-the-fly~\cite{waters2019leveraging}. Also, recent studies~\cite{wang2019signal,chandak2020streaming} demonstrate that language identification (LID) can be improved by combining conventional acoustic representations with textual cues from the ASR decoder. To leverage these ideas---and also the fact that RNN-T models the joint evolution of acoustic and lexical information---, we propose a multilingual RNN-T architecture that consumes embeddings from an acoustic LID classifier and predicts both the spoken content and the spoken language. Accurate LID can improve ASR performance via language-specific second-pass rescoring and also enable appropriate downstream processing by language-specific components like natural language understanding. 
Furthermore, multilingual joint ASR-LID models can simplify dynamic language switching by replacing several monolingual ASRs and a language detector with just one system. 
The key contributions of this paper are summarized below.

$\bullet$ Joint ASR-LID: We append ground-truth transcripts with language tags and include them in the model's vocabulary. This approach has been explored for LAS~\cite{watanabe2017language}, but, to the best of our knowledge, has not been studied in the context of RNN-T.

$\bullet$ Acoustic LID embeddings: Frame-level embeddings from an acoustic LID classifier (trained beforehand) are used to guide RNN-T training and inference. This is similar in principle to the idea of deriving language information from an auxiliary RNN-T \cite{waters2019leveraging}, but it has certain advantages (see Section~\ref{sec:methods}).

$\bullet$ Biasing joint network: Previous studies provide language information to RNN-T's encoder or decoder~\cite{kannan2019large}. In this work, we study the effect of influencing RNN-T's joint network.

$\bullet$ Languages: Previous studies mostly focus on groups of related languages, e.g. Indic, Arabic and Nordic~\cite{kannan2019large, waters2019leveraging}. 
Motivated by the real-world usage patterns in bilingual markets, this work explores two pairs of unrelated languages: English-Hindi and English-Spanish.

In the rest of this paper, \textit{multilingual} and \textit{bilingual} will be used interchangeably---the latter will mostly be used when referring to experimental details or results.

\section{Methods}\label{sec:methods}

A typical RNN-T ASR architecture comprises a transcription network (encoder), a prediction network (decoder) and a joint network. The encoder is an RNN that converts an input acoustic feature vector, $\mathbf{x}_t$, to a hidden representation, $\mathbf{h}^{enc}_t$, and the decoder is an RNN that receives the last non-blank label observed ($y_{u-1}$) and outputs a hidden representation, $\mathbf{h}^{dec}_u$:
\begin{equation}
\mathbf{h}^{enc}_t = f^{enc}(\mathbf{x}_t), \text{  } \mathbf{h}^{dec}_u = f^{dec}(y_{u-1}).
\end{equation}
The joint network is a feedforward network that combines the encoder and decoder outputs to produce logits, $\mathbf{z}_{t,u}$:
\begin{equation}
\mathbf{z}_{t,u} =  f^{joint}(\mathbf{h}^{enc}_t, \mathbf{h}^{dec}_u),
\end{equation}
which in turn are passed through a softmax layer to produce a probability distribution for the next output symbol, $y_u$ (either blank or one of the ASR targets):
\begin{equation}
P(y_u|t,u) = \mathrm{softmax}(\mathbf{z}_{t,u}).
\end{equation}
A naive approach to train multilingual ASR using RNN-T is to simply pool data from the languages of interest and define the output symbol space as the union of the individual symbol sets. As literature has shown, better results could be achieved by supplementing $\mathbf{x}_t$ with auxiliary language information, $\mathbf{l}_t$, to obtain an improved encoder representation, $\mathbf{g}^{enc}_t$:
\begin{equation}
\mathbf{g}^{enc}_t = f^{enc}([\mathbf{x}_t; \mathbf{l}_t]).
\end{equation}
In this paper, we also study the effect of supplying language information to the joint network. This enables us to determine if language information is more helpful at the input where lower-level features are extracted, or deeper inside the network where higher-level acoustic and lexical information are combined. Depending on whether $\mathbf{l}_t$ is provided to the encoder, joint network or both, the model's logits can be computed using Eq.~\eqref{eq:lidenc}, \eqref{eq:lidjnt} or \eqref{eq:lidboth}, respectively. Further implementation details are presented in Sections~\ref{subsec:oracle} and \ref{subsec:inferred}.
\begin{eqnarray}
\mathbf{z}^E_{t,u} &=&  f^{joint}(\mathbf{g}^{enc}_t, \mathbf{h}^{dec}_u), \label{eq:lidenc} \\
\mathbf{z}^J_{t,u} &=& f^{joint}(\mathbf{h}^{enc}_t, \mathbf{l}_t, \mathbf{h}^{dec}_u), \label{eq:lidjnt} \\
\mathbf{z}^B_{t,u} &=& f^{joint}(\mathbf{g}^{enc}_t, \mathbf{l}_t, \mathbf{h}^{dec}_u). \label{eq:lidboth}
\end{eqnarray}

\begin{figure}[t]
  \centering
  \includegraphics[width=\linewidth]{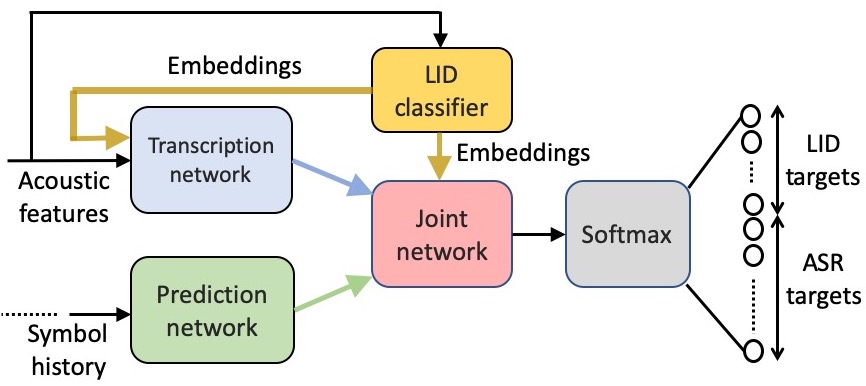}
  \caption{A schematic representation of the proposed RNN-T based multilingual joint ASR-LID model.}
  \label{fig:block_diagram}
  \vspace{-0.1in}
\end{figure}

To train a multilingual joint ASR-LID model we extend the RNN-T's output symbol space to include language targets. Further details related to training and inference are presented in Section~\ref{subsec:jtrn}.
The proposed multilingual joint ASR-LID architecture is schematically summarized in Fig.~\ref{fig:block_diagram}.

\subsection{Baselines}\label{subsec:bsl}

ASR performance is evaluated against two baselines: monolingual RNN-T models (\texttt{A0}) and multilingual RNN-T models trained without utilizing language adaptation methods, i.e. by simply pooling training data from the languages of interest (\texttt{A1}).
LID performance is evaluated against an acoustic-only baseline, i.e. a recurrent architecture trained to classify the languages of interest using acoustic features only.

\subsection{Multilingual ASR using oracle language identity}\label{subsec:oracle}

The simple pooling approach (\texttt{A1}) is known to be suboptimal, and several studies have shown that language-adaptive training is essential to improving the recognition accuracy of multilingual ASR \cite{mueller2015,muller2019neural,kannan2019large}. A popular way of achieving this is to encode the language identity as a constant (utterance-level) \textit{one-hot} vector. Since language identity is unknown at runtime in dynamic multilingual settings, we treat one-hot language encoding as an \textit{oracle} approach.
As mentioned earlier (Eqs.~\eqref{eq:lidenc}--\eqref{eq:lidboth}), we evaluate the effect of supplying one-hot language vectors to different parts of the RNN-T: encoder input (\texttt{A2}$^{E}$), joint network (\texttt{A2}$^{J}$), or both (\texttt{A2}$^{B}$).

\subsection{Multilingual ASR using inferred language identity}\label{subsec:inferred}

To achieve realistic language-adaptive training, the one-hot vectors in \texttt{A2} must be replaced with inferred language information. In \cite{waters2019leveraging}, an auxiliary RNN-T model is used to derive language information on-the-fly (a sequence of blanks until the language tag is emitted followed by another sequence of blanks); while this approach is effective, an auxiliary RNN-T must be trained for the same language group that the ASR system is designed to recognize. In this paper, we propose to use frame-wise embeddings from an acoustic LID classifier that is trained beforehand. If the LID classifier is trained on a large group of languages, one could potentially use the same model to generate embeddings regardless of the ASR's target language group. 
Another advantage of using acoustic embeddings is that they can be easily consumed by the joint network to be combined with the encoder and decoder representations.
Note however that the guidance that the ASR network receives early on in an utterance may not be very valuable as the language embeddings are updated on a frame-by-frame basis, and there may not be enough evidence to accurately detect the language after observing just a few frames of audio. Similar to the oracle experiments, we evaluate three different ways of supplying inferred language information to the RNN-T (\texttt{A3}$^{E}$, \texttt{A3}$^{J}$ and \texttt{A3}$^{B}$). 

\subsection{Multilingual joint ASR-LID modeling}\label{subsec:jtrn}

To train multilingual joint ASR-LID models, we extend the RNN-T's output symbol space to include language targets---meaning that the model can output either blank, one of the language tags, or one of the ASR tokens. Training targets are generated by simply appending language tags to the ground truth transcripts. The intuition behind having language tags as the utterance-final tokens is that the network's belief of the underlying language improves with the incoming acoustic feature frames and the partially predicted text. Akin to the end of sentence or \texttt{<eos>}  symbol—used to indicate end of utterance and optionally perform endpointing—, appending language tags to utterance ends encourages the network to make language predictions after sufficient evidence has been observed.

\begin{table}[t]
\begin{center}
\caption{Dataset sizes in hours. ASR and LID training data are used for RNN-T and acoustic LID training, respectively.}\label{table:datasets}
\begin{tabular}{c|cc|c}
\hline
\textbf{System; Language} & \multicolumn{2}{c|}{\textbf{Train}} & \textbf{Test} \\
\hline
\multicolumn{1}{c}{} & ASR & \multicolumn{1}{c}{LID} & ASR, LID \\ 
\hline
English-Spanish; en-us & 6.2k & 1.9k & 22 \\
English-Spanish; esp-us & 3.5k & 1.7k & 22 \\ \hline
English-Hindi; en-in & 12.6k & 1.0k & 83 \\
English-Hindi; hi-in & 2.3k & 1.1k & 24 \\
\hline
\end{tabular}
\end{center}
\vspace{-0.3in}
\end{table}

Joint ASR-LID models can be trained without (\texttt{A4}) or with (\texttt{A5}$^{E}$, \texttt{A5}$^{J}$, and \texttt{A5}$^{B}$) language embeddings on the input side.
Based on the observations made for \texttt{<eos>}-driven endpointing using RNN-T \cite{chang2019joint}, ASR deletions are expected to increase with the introduction of auxiliary language tags.
Therefore, inspired by \cite{chang2019joint}, we employ emission penalties during beam search. For language tags to be considered valid candidates in the decoding beam, language posteriors must exceed a threshold $\beta$ after being modified by an exponent parameter $\alpha$ ($\alpha > 1$ implies higher penalty). Eq.~\eqref{Eq:alpha_beta} summarizes this condition.
\begin{equation}
\label{Eq:alpha_beta}
P(y_{u}| t, u )^ \alpha >= \beta  \text{  if  } y_{u}  \in  \{\text{language targets}\} 
\end{equation}

We evaluate ASR and LID performance for different values of $\alpha$ and $\beta$, including the extreme case of $\beta=1$ which prevents the model from emitting language tags altogether. While the penalized inference mechanism---which suppresses premature language tag emissions---is expected to minimize deletions and thereby improve ASR performance, LID performance might suffer if predictions are made based on the model's 1-best output. Therefore, in all our experiments with joint models, language is always predicted using the unscaled model posteriors available after the last audio frame; in other words, language prediction does not rely on the presence of language tags in the 1-best output. Note that language tags, when emitted, are also ignored during word error rate (WER) computation.

\section{Experimental setup}\label{sec:expsetup}

\begin{table}[htb!]
\begin{center}
\caption{A summary of the experimental setup.}\label{table:expsetup}
\begin{tabularx}{\linewidth}{X}
\hline
\multicolumn{1}{c}{\textbf{Acoustic features}} \\
\hline
$\bullet$ 64 dimensional log filter-bank energies (LFBEs) extracted at 10 ms intervals using 25 ms windows $\rightarrow$ Feature normalization $\rightarrow$ three frame stacking $\rightarrow$ downsampled to 30 ms \\
$\bullet$ SpecAugment with frequency masking \cite{park2019specaugment} employed for RNN-T training; two masks applied per utterance with a maximum width of 24 channels \\
$\bullet$  Acoustic LID: augmentation via simulated reverberation \\
\hline
\multicolumn{1}{c}{\textbf{Training targets}} \\
\hline
$\bullet$ RNN-T: vocabulary of 4k subwords via byte pair encoding \cite{sennrich2015neural}, using language-specific and pooled training text corpora respectively for monolingual and bilingual models \\
$\bullet$ Acoustic LID: utterance-level language labels \\
\hline
\multicolumn{1}{c}{\textbf{Model architectures}} \\
\hline
$\bullet$ RNN-T uses 5 encoder LSTM layers and 2 decoder LSTM layers with 1024 units each and a 512 dimensional embedding layer at decoder input; joint network has 512 hidden units followed by $\tanh$ and softmax \\
$\bullet$ Acoustic LID uses 3 LSTM layers with 256 units each, followed by 32 dimensional projection and softmax; output of projection layer provided as embeddings to RNN-T \\
\hline
\multicolumn{1}{c}{\textbf{Training}} \\
\hline
$\bullet$ Distributed training on 24 and 16 GPUs respectively for RNN-T and acoustic LID; batch size of 64 per GPU \\
$\bullet$ Dropout rate of 0.2 used in RNN-T encoder and decoder \\
$\bullet$ Adam optimizer used with a warmup-hold-decay learning rate (LR) schedule \\
$\bullet$ Stratified sampling: training batches retain the natural language distribution of the pooled dataset \\
$\bullet$  All RNN-T models trained for 225k steps \\
\hline
\multicolumn{1}{c}{\textbf{Inference}} \\
\hline
$\bullet$ ASR 1-best obtained using a beam width of 16 and a temperature of 1 \\
$\bullet$ Acoustic LID models after 75k steps of training used for embedding extraction and baseline evaluation \\
\hline
\end{tabularx}
\end{center}
\vspace{-0.3in}
\end{table}

\begin{table*}[t]
\begin{center}
\caption{ASR WERRs \underline{relative} to monolingual ASR models and LID accuracies \underline{relative} to acoustic-only LID classifiers; higher numbers imply better performance in both cases. $E$, $J$ and $B$ carry the same meaning as in Eqs.~\eqref{eq:lidenc}--\eqref{eq:lidboth}. For \texttt{A7}, WW = ``wake-word".}
\label{table:werr_table}
\begin{tabular}{c|c|c|cc|cc|cc|cc}
\hline
\hline

\textbf{\multirow{3}{*}{Model}} & \textbf{\multirow{3}{*}{Description}}  & \textbf{\multirow{3}{*}{($\alpha,\beta$)}} & \multicolumn{4}{c|} {\textbf{Relative ASR WERR ($\%$)}} & \multicolumn{4}{c}{\textbf{Relative LID Accuracy ($\%$)}} \\
\cline{4-11}
 &  & & \multicolumn{2}{c|}{English-Hindi} & \multicolumn{2}{c|}{English-Spanish} & \multicolumn{2}{c|}{English-Hindi} & \multicolumn{2}{c}{English-Spanish}  \\
 \cline{4-11}
    &        &  & en-in & hi-in & en-us & esp-us & en-in & hi-in & en-us & esp-us \\
     \hline
 & Acoustic LID classifier   &  -- &    --        & -- & -- &    --  & 0.0 & 0.0 & 0.0 & 0.0 \\ \hline
\texttt{A0} & Monolingual training    & --                  & 0.0 & 0.0 & 0.0 & 0.0& -- & -- & -- & -- \\ \hline

\texttt{A1} & Simple data pooling    & --         & -9.4 & -11.2 & -0.5 & 	0.2 & -- & -- & -- & --\\
\hline

\texttt{A2}$^{E}$ & \multirow{3}{*}{Oracle language input} & -- &-1.9 & 13.1 & -0.2 & 6.3 & -- & -- & -- & -- \\
\texttt{A2}$^{J}$ &  & -- & -1.8	 &  10.6	 &  0.0 & 4.4 & -- & -- & -- & --\\
\texttt{A2}$^{B}$ & & -- & -1.3 & 14.7 & 0.0 & 4.1 & -- & -- & -- & --\\
\hline
\texttt{A3}$^{E}$ & \multirow{3}{*}{Inferred LID embeddings} & -- & -9.4 & -20.3 &  -2.5 & 0.1 & -- & -- & -- & --\\
\texttt{A3}$^{J}$ &    & -- & -8.3 &  -19.1 & -0.4 & 0.0 & -- & -- & -- & --\\
\texttt{A3}$^{B}$ &  & -- & -8.9 & -21.0	 & -3.4 & -0.8 & -- & -- & -- & --\\
\hline
\texttt{A4} & \multirow{3}{*}{Joint training} & (1, 0)  & -64.6 &  -18.5 & -57.4 & -69.8  &-11.7 &  -0.4 & 1.3 &  -0.1\\
\texttt{A4} & & (2, 0.1)  & -27.8	 & -4.5 & -21.8 & -25.1 & -11.9 &  -0.5 & 1.3 & 0.2\\
\texttt{A4} & & (1, 1) & -22.4 & -1.8 &  -7.2 & -5 & -21.1 & -1.4	& 1.7 & -3.1 \\
\hline
\texttt{A5}$^{E}$ & Inferred LID embeddings &  \multirow{3}{*}{(1, 1)} & -14.3  &  0.7	 & 3.5  & -2.7 & -5.4 & -2.4 &   -2.7 & 0.3 \\
\texttt{A5}$^{J}$ & + & & -6.8 & 4.9 &  -1.0 & -1.6 & -7.4 & -1.6 & -3.8	 &  1.1 \\
\texttt{A5}$^{B}$ & Joint training &  & -18.7 & 0.3	 & -2.3 & -2.0 & -6.9  & -1.9   & 0.8 & -0.1 \\
\hline

\texttt{A6}  &\texttt{A5}$^{J}$ with LID posteriors & (1, 1) & -7.0 & 4.4 &  -0.7 & 0.8 & -6.1 & -1.6 & 0.1	 &  0.3 \\ \hline
\texttt{A7}  &\texttt{A6} without WW-only utterances & (1, 1) & -4.9 & 4.7 & -0.2 & 0.9 & -5.5  & -1.2   & -2.1 & 1.4 \\

\hline
\hline
\end{tabular}
\end{center}
\vspace{-0.2in}
\end{table*}


We conduct experiments using two pairs of languages: English-Spanish, as spoken in the United States (languages denoted as en-us and esp-us, respectively), and English-Hindi, as spoken in India (languages denoted as en-in and hi-in, respectively). Table~\ref{table:datasets} captures the dataset statistics. 

Of the two language pairs studied, English-Hindi shows a higher degree of within-utterance code switching owing to the colloquial nature of spoken Hindi (which involves frequent use of common English words such as ``play", ``call", ``book", ``volume", etc.) and the presence of named entities in a number of other Indian languages. Utterances in en-in are transcribed using Latin script, whereas utterances in hi-in are transcribed using both Latin and Devanagari scripts: the latter for Hindi words and the former for words in English and other languages. 
Since named entities such as Hindi movie titles, song names, etc. are common to both the languages, a bilingual model can probabilistically output several words in either Latin or Devanagari. These aspects make English-Hindi a challenging language pair to work with.
Table~\ref{table:expsetup} summarizes the remaining details pertaining to our experimental setup.

\section{Results and discussion}

Table \ref{table:werr_table} summarizes the results obtained from all of our experiments. We report word error rate reductions (WERR) relative to monolingual ASR systems, and LID accuracies relative to acoustic LID classifiers.

\subsection{Results using data pooling and oracle language inputs}
The simple data pooling approach (\texttt{A1}) results in performance degradations in general. This is more prominent for the English-Hindi language pair, where a large number of common words have Latin and Devanagari representations, respectively, in the two languages which can cause potential script-based confusions with pooled training data. Providing oracle language information (\texttt{A2}$^{E}$, \texttt{A2}$^{J}$, \texttt{A2}$^{B}$) yields healthy WERRs, especially for hi-in and esp-us, the relatively underrepresented languages in each pair. These results validate the benefits of incorporating language information, and serve as an upper bound for the gains achievable with language based input.

\subsection{Results using inferred language embeddings}
Next we evaluate the models trained by incorporating on-the-fly LID embeddings from a pretrained acoustic LID classifier. The performance for the (en-us, esp-us) pair, with \texttt{A3}$^{J}$, seems to be almost on par with the baseline.
For hi-in however, a significant drop in performance is observed. One plausible explanation is the presence of Indic words in English utterances, and significant code switching in the Hindi utterances, which reduces the discriminatory power of the embeddings. Note that we use frame-level language embeddings in order to build a streaming-friendly model, and hence the model could be susceptible to the within-utterance language fluctuations.

\subsection{Results for joint ASR-LID training}

For the joint training approach (\texttt{A4}) with vanilla inference strategy $(\alpha=1,\beta=0)$, we observe a dramatic increase in WERs (large negative WERRs) across languages, with the increase in deletions (owing to early language tag emissions) being the largest contributor (2--6 times higher deletions observed as compared to the monolingual counterparts). By regulating language tag emissions via $\alpha$ and $\beta$, we observe that the ASR metrics improve. The extreme penalty scheme $(\beta=1)$, where language tags are never emitted, performs best in terms of WERR. Note that this choice doesn't impact LID accuracy significantly since that is governed solely by utterance-final posteriors.

Combining joint training with inferred LID embeddings (\texttt{A5}$^{E}$, \texttt{A5}$^{J}$, \texttt{A5}$^{B}$) brings us one step closer to bridging the performance gap relative to the ASR and LID baselines. \texttt{A5}$^{J}$ offers the best results in general, and its performance for (en-us, esp-us) is comparable to or slightly worse than that of the baseline systems.
Results for the more difficult (en-in, hi-in) pair show interesting trends. For hi-in, \texttt{A5}$^{J}$ yields a 4.9\% WERR and an LID accuracy that is slightly worse than the baseline, while for en-in, both ASR and LID metrics show larger gaps relative to their respective baselines (although the gaps are much smaller as compared to \texttt{A3}$^{J}$ and \texttt{A4}).
These results suggest that the embedding-driven joint models might have learnt to become more consistent---but not fully accurate---with regard to the output script produced. 

We improve upon \texttt{A5}$^{J}$ in two ways. (1) Posteriors from the acoustic LID model are used instead of embeddings from its projection layer in order to provide a more discriminative signal to the network (similar to the oracle language input). This approach (\texttt{A6}) yields marginal performance gains. (2) Since a significant portion of Alexa traffic comprises wake-word only utterances (``alexa"), models could be confused between accent and language (e.g., an en-us ``alexa” utterance could be interpreted as esp-us if it has a strong Spanish accent, which is undesirable). To mitigate such effects, we train \texttt{A6} after filtering out all utterances with wake-words only and observe useful improvements in general (\texttt{A7}). 

\section{Conclusions}

This paper employs the RNN-T architecture to build streaming, end-to-end, bilingual systems for joint speech recognition and language identification. We use a lightweight acoustic LID classifier to provide on-the-fly language embeddings to different components of the RNN-T, and demonstrate that providing language information to the joint network performs best. 
By penalizing language tag emissions and making language predictions using utterance-final posteriors, we show that ASR performance can be improved without impacting LID accuracy.
The modeling techniques proposed in this work are language agnostic and can be scaled to multiple languages.

For the English-Spanish language pair, the joint ASR-LID model achieves comparable performance relative to the monolingual ASR and acoustic LID systems. In the case of English-Hindi, we observe a slight performance degradation for English owing to code switching effects combined with the presence of dual (Latin and Devanagari) word representations. For Hindi, the joint model surpasses baseline ASR performance while almost matching LID accuracy. The experimental evidence from two significantly different language pairs indicates that joint ASR-LID training is a promising direction to pursue, given the modeling simplification and compute savings it offers.

\bibliographystyle{IEEEtran}

\bibliography{joint_asr_lid}

\end{document}